Costantino Sigismondi

# MISURA DEL RITARDO ACCUMULATO DALLA ROTAZIONE TERRESTRE, ΔUT1, ALLA MERIDIANA CLEMENTINA DELLA BASILICA DI SANTA MARIA DEGLI ANGELI IN ROMA

Abstract. *The Clementine Gnomon is a solar meridian telescope dedicated to solar astrometry operating as a giant pinhole dark camera, being the basilica of Santa Maria degli Angeli the dark room. This instrument built in 1701-1702 by the will of pope Clement XI by Francesco Bianchini (1662-1729) gives solar images free from distortions, excepted atmospheric refraction, because the pinhole is opticsless. Similar historical instruments are in Florence (Duomo, by Toscanelli and Ximenes), Bologna (San Petronio, by Cassini), Milan (Duomo, by De Cesaris) and Palermo (Cathedral, by Piazzi). The azimut of the Clementine Gnomon has been recently referenced with respect to the celestial North pole, and it is 4′28.8″±0.6″, a comparison with similar coeval instruments is presented. Also the local deviations from a perfect line are known with an accuracy better than 0.5 mm. With these calibration data we used the Gnomon to measure the delay of the solar meridian transit with respect to the time calculated by the ephemerides (ΔUT1). The growth of this astronomical parameter is compensated by the insertion of a leap second ad the end of the year in order to keep the Universal Time close to astronomical phenomena within less than a whole second. On December 31, 2008 at 23:59:59 there is one of those leap seconds leading to 23:59:60 before the new year's midnight 00:00:00, being ΔUT1≈0.7 s at that date; the last similar event occurred on December 31, 2005. There are several modern observatories dedicated to astrometry which contribute to measure daily ΔUT1 under the coordination of* Institute of Earth rotation and Reference system Service IERS, *and we show how it is possible to perform such a measurement with the Clementine Gnomon. We emphasize the opportunity of considering the Clementine Gnomon as introductory in modern astrometry besides its key role in the history of astronomy. The need of a definitive solution in restoring the original pinhole is also shown. The adopted technique is videorecording at 60 fps solar transits with absolute timing reference. The reference ephemerides are those from the* Institute de Mécanique Céleste et Calcul des Ephémérides IMCCE *and those of the Astronomical Almanac. ΔUT1 is measured with an accuracy of ±0.3 s.*

## 1. *Lo Gnomone Clementino*

Lo Gnomone Clementino è un telescopio meridiano a camera oscura, che si trova nella basilica di Santa Maria degli Angeli a Roma[1]. Lo strumento fu voluto e finanziato da papa Clemente XI Gianfrancesco Albani (1700-

[1] Bianchini (1703).

1721), tra i primi atti del suo pontificato, proprio nella chiesa dove egli, appena ordinato sacerdote, aveva celebrato la sua prima messa solenne il 6 ottobre 1700[2]. L'astronomo e archeologo Francesco Bianchini (1662-1729), veronese, allievo di Geminiano Montanari a Padova, fu l'artefice di questo strumento di dimensioni colossali: 20 metri di altezza e quasi 50 di lunghezza, realizzato con marmi preziosissimi[3].

La scelta della basilica costruita da Michelangelo sulle mura romane delle Terme di Diocleziano risolveva anche il problema che Giandomenico Cassini aveva riscontrato a S. Petronio a Bologna, dove la quota del foro stenopeico era variata leggermente in 40 anni a causa degli assestamenti dell'edificio[4]. I medesimi problemi furono poi riscontrati dal gesuita Leonardo Ximenes allo gnomone di Toscanelli (1475) del duomo di Firenze[5].

## 2. *L'azimut delle meridiane*

La ricognizione della meridiana del Duomo di Milano del 1977[6], è stata eseguita con i metodi della moderna geodesia usando la stella Polare[7], applicati da noi nel 2006 anche allo Gnomone Clementino[8].

Il metodo dell'osservazione dei passaggi della Polare col teodolite consiste nelle seguenti operazioni.

Si fissa una linea poligonale di cui fanno parte gli estremi della meridiana ed una linea ausiliaria all'esterno della chiesa. Si misurano con il teodolite tutti gli angoli della poligonale. Si punta la stella Polare e se ne osserva il lento moto (la velocità angolare massima è 0.12"/s) attraverso il cannocchiale del teodolite. Se la stella viene sfuocata si può osservare un transito della macchia di luce sulla croce del mirino e valutare l'istante in cui la croce divide a metà la macchia. Si trascrive l'istante letto su un orologio sincronizzato con il Tempo universale coordinato UTC, quello trasmesso dall'Istituto elettrotecnico Galileo Ferraris di Torino sulle frequenze di Radio1 nelle ore intere. Con un programma di effemeridi si ricava l'azimut della Polare a quel determinato istante. Un secondo di precisione nella misura del tempo è sufficiente a dare 0.12" di precisione in azimut. Si trascrive anche l'angolo che la direzione del teodolite forma con la linea esterna ausiliaria. Si ripete la misura con il teodolite ruotato di 180° in azimut e quanto necessario in altezza per rivedere la Polare (metodo di Bessel per compensare gli errori di allineamento degli assi del cannocchiale), e si fa la media tra la misura diretta e quella a 180°. Sono sufficienti 3 serie di queste misure per avere già una precisione assoluta inferiore al secondo d'arco, come si può vedere dalla dispersione delle misure ottenute.

[2] F. Bianchini, *Carteggi*, conservati presso la Biblioteca Vallicelliana Roma.
[3] Catamo – Lucarini (2002).
[4] Paltrinieri (2001).
[5] Bònoli – Parmeggiani – Poppi (2006).
[6] Ferrari da Passano – Monti – Mussio (1977).
[7] Bezoari – Monti – Selvini (2002).
[8] Sigismondi (2008a).

Usando l'immagine del Sole in chiesa, anziché la Polare, si ha che la velocità angolare del Sole è 15″/s, per cui occorre l'analisi dei singoli fotogrammi del video (1/60 s per le misure qui presentate) per avvicinarsi alla precisione del metodo della Polare senza dover utilizzare il teodolite e la linea esterna ausiliaria.

Si noti che l'azimut da referenziare è definito come quello della retta passante per i punti estremi della meridiana: il punto più a nord ed il piede del foro stenopeico. Localmente, giorno dopo giorno, l'immagine del Sole passa su parti della linea reale che possono essere leggermente disallineate rispetto a questa retta ideale, questi disallineamenti sono generalmente piccoli, qualche millimetro, ma capaci di modificare la misura dell'azimut di svariati secondi d'arco.

Per lo Gnomone Clementino abbiamo ottenuto i seguenti valori dello scarto dalla retta.

FIG. 17.1. *Deviazioni locali dalla retta che va dal piede della perpendicolare del foro all'estremo Nord della linea meridiana. L'unità di misura in ascisse vale 203.44 mm ed è la centesima parte dell'altezza del foro stenopeico. In ordinate abbiamo i millimetri, contati col segno positivo quando la linea reale si trova ad ovest della retta, negativo quando è est. Si vede chiaramente il grande "seno" menzionato dal gesuita Boscovich (1773) autore di una accurata ricognizione dello strumento nel 1750*[9]*. Per le misure qui riportate abbiamo materializzato la retta con un raggio laser passante per gli estremi della meridiana*

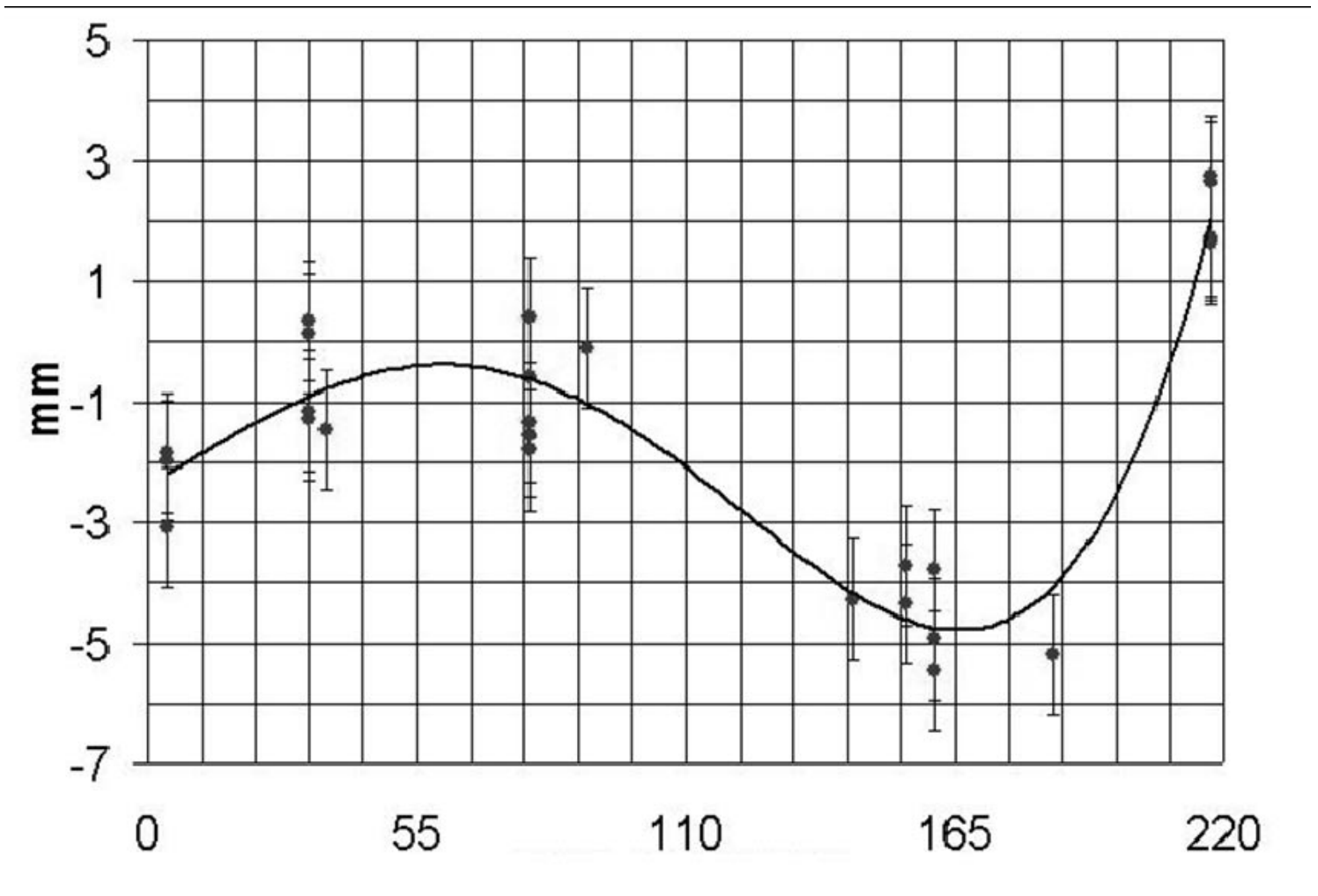

[9] Boscovich, Lettera a A. Vallisneri del 25 agosto 1773 in PAOLI (1988, p. 283).

TAB. 17.1. *Azimut delle grandi meridiane. Eccetto quella di Tycho Brahe sono tutte ancora funzionanti*

| | |
|---|---|
| Toscanelli 1475 (Duomo Firenze)[a] | -19′ 54″ |
| Tycho Brahe 1571 (Uraniborg, Hven)[10] | -17′ |
| Cassini 1655 (San Petronio, Bologna)[11] | +1′ 36.6″ |
| Bianchini 1702 (Santa Maria degli Angeli) | +4′ 28.8″±0.6″ |
| Celsius 1734 (Santa Maria degli Angeli)[12] | +2′ |
| Boscovich e Maire 1750 (S. M. degli Angeli)[13] | +4′ 30″ |
| Ximenes 1761 (Duomo Firenze)[a] | -27″ |
| De Cesaris 1786 (Duomo Milano)[14] | +0′ 07″ |
| Gigli 1817(Obelisco di Piazza S. Pietro, Roma)[a] | -5′36″ |

[a] Misure in loco di Costantino Sigismondi con la tecnica del confronto tra video ed effemeridi qui descritta.
Le misure a Firenze sono state svolte con il solo video sincronizzato UTC, senza tenere in conto eventuali deviazioni locali della linea, o modifiche della posizione del piede della verticale del foro rispetto alla configurazione originale non più misurabile sia per lo gnomone di Toscanelli che per quello dello Ximenes.

Le ragioni di queste differenze dal vero nord non sono ancora chiare, probabilmente legate alle effemeridi del tempo, i cui metodi di calcolo andavano via via sempre migliorando.

Sia Celsius che i gesuiti Boscovich e Maire effettuarono le loro ricognizioni sullo Gnomone Clementino con una loro meridiana di riferimento più piccola. Celsius la aveva al Quirinale, i gesuiti al Collegio Romano.

La meridiana veniva allineata con il nord partendo da principi primi: mi riferisco al passaggio del Sole attraverso cerchi di uguale altezza che avviene, attorno ai solstizi, ad azimut simmetrici rispetto al sud.

Una volta realizzato il foro stenopeico e disegnato un opportuno cerchio centrato nel piede della verticale per il foro, si prende il punto medio della retta che congiunge i centri delle immagini del Sole mentre attraversavano quel cerchio al mattino e al pomeriggio. Congiungendo questo punto al centro del cerchio si ottiene la linea meridiana.

[10] *Ibid*.
[11] PALTRINIERI (2001).
[12] MONACO (1990, pp. 118-119).
[13] Boscovich, Lettera a A. Vallisneri del 25 agosto 1773 in PAOLI (1988, p. 283); HEILBRON (2005, p. 216).
[14] FERRARI DA PASSANO – MONTI – MUSSIO (1977, p. 37).

È probabile che al diminuire delle dimensioni delle meridiane di riferimento l'errore angolare, casuale, aumenti, ma Boscovich (1759) a proposito della meridiana di Uraniborg commentava «uno sbaglio enorme non può in conto alcuno attribuirsi all'incuria di Ticone, di cui si sa bene la diligenza e la grandezza, e la bontà degl'istrumenti, che adoperava inferiori veramente a nostri, ma incomparabilmente migliori di quello che si richiede per evitare un tanto errore»[15].

### 3. *L'Istante del Transito Meridiano*

L'obiettivo dello Gnomone è un foro stenopeico, dunque è privo di elementi ottici che distorcerebbero necessariamente le immagini del campo di vista. L'unica causa di distorsione è la rifrazione atmosferica comune a tutti gli osservatori sulla Terra, che però non influenza gli istanti di contatto dei lembi solari con la linea meridiana. Facendo la media delle misure dei due tempi di contatto dei lembi dell'immagine stenopeica del Sole con la linea meridiana si ottiene l'istante del transito del Sole sulla meridiana con una precisione di ±0.5 s ad occhio nudo, e migliore di ±0.1 s con una videocamera sincronizzata col Tempo universale coordinato UTC e la tecnica dei transiti paralleli[16].

FIG. 17.2. *Schema del passaggio dell'immagine solare sulla linea meridiana, con gli istanti di contatto*

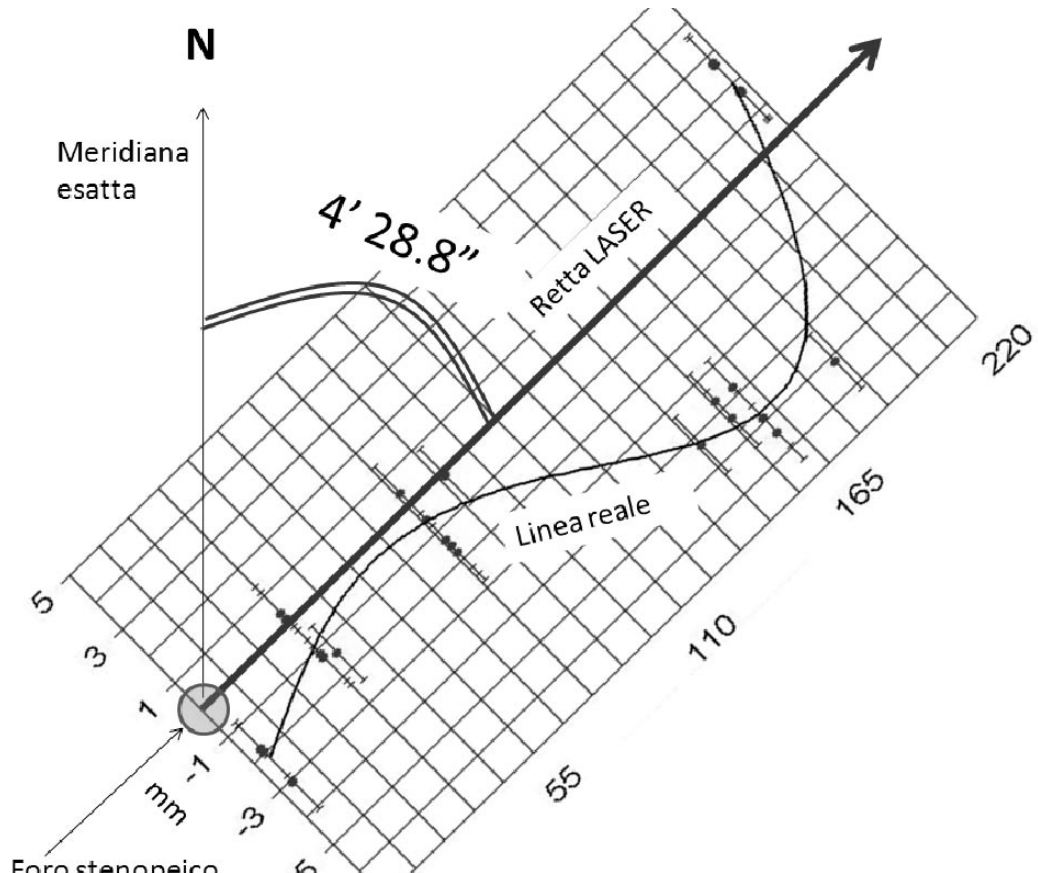

[15] Boscovich, Lettera a A. Vallisneri del 25 agosto 1773 in PAOLI (1988, p. 283).
[16] SIGISMONDI (2006).

Le effemeridi, dal greco *ephemeron*, che significa giornaliero, sono tabelle di grandezze astronomiche calcolate per ogni istante voluto. Oggi la maggior parte sono disponibili via web. Noi utilizziamo quelle del Sole sull'*Astronomical Almanac*[17] pubblicato dall'*US Naval Observatory*, dell'*IMCCE*, l'*Institute de Mécanique Céleste et Calcul des Ephémérides*, sul sito www.imcce.fr, e del *NREL, National Renewable Energy Laboratory* (SPA, *Solar Position Algoritm*[18]), e Ephemvga[19].

FIG. 17.3. *Schema per distinguere la meridiana esatta dalla retta LASER e dalla linea reale*

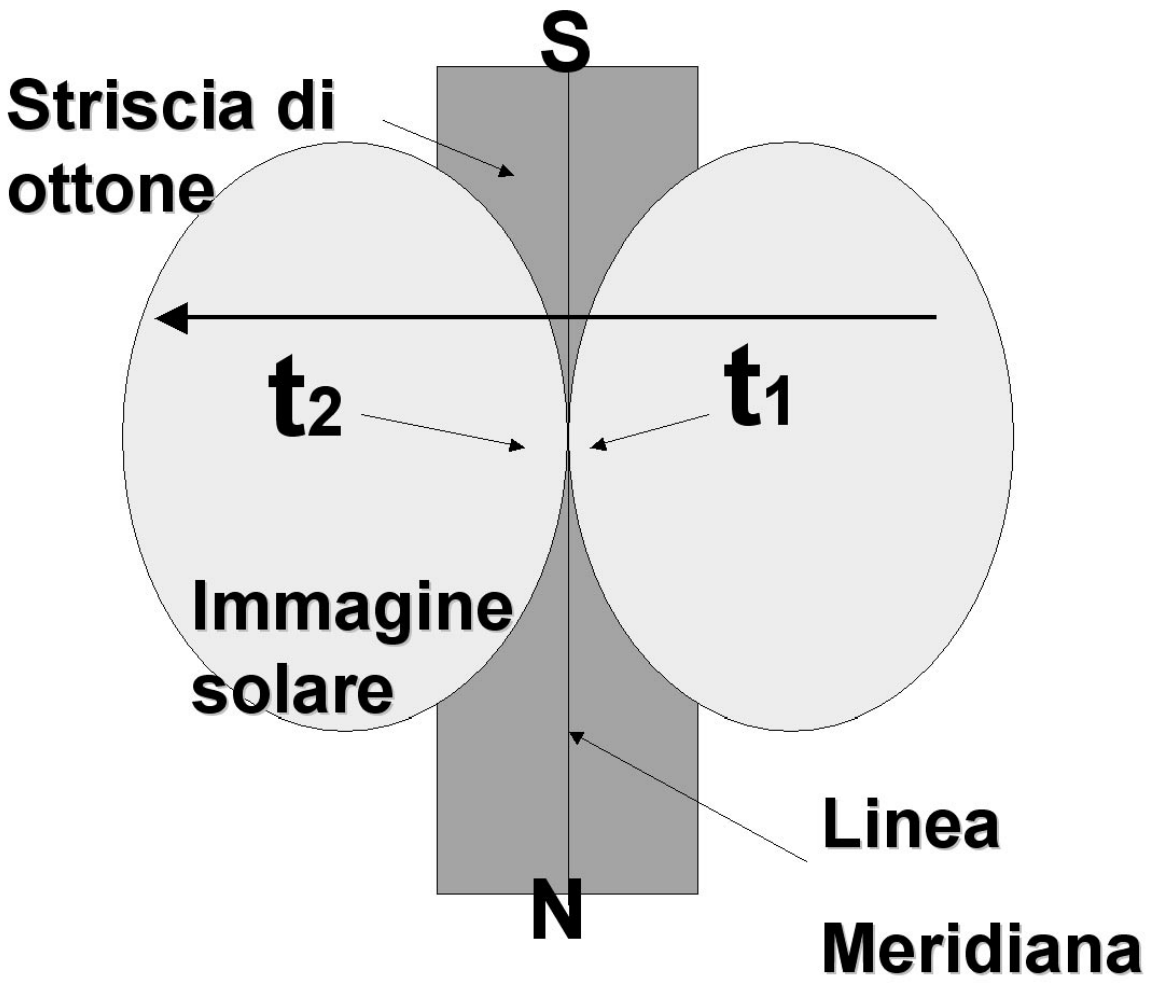

Per poter fare delle misure astrometriche con lo Gnomone Clementino dobbiamo riportare le misure prese sulla Linea meridiana "reale" a quella esatta, diretta precisamente lungo il meridiano locale nord-sud.

Per poter svolgere questa operazione dobbiamo conoscere l'azimut della retta LASER e le deviazioni locali dalla retta LASER della linea reale.

4. *La misura del ΔUT1*

Ad esempio il 3 dicembre 2008 il centro del Sole è transitato sulla linea reale al punto 205.5 dello Gnomone Clementino alle 12:00:15.65 UTC. Le effemeridi IMCCE prevedono il passaggio al meridiano esatto alle 11:59:57.87,

[17] US NAVAL OBSERVATORY HM NAUTICAL ALMANAC OFFICE (2008).
[18] http://www.nrel.gov/midc/solpos/spa.html.
[19] www.santamariadegliangeliroma.it menu meridiana – calcolo delle effemeridi.

17.78 s prima. Localmente la linea reale si trova 1.35 mm a ovest della retta LASER.

La velocità dell'immagine solare, diretta perpendicolarmente alla linea da ovest verso est, è per quel punto 3.13 mm/s, a tale velocità la deviazione locale si copre in 0.43 s. Dunque il ritardo complessivo misurato in due tempi (osservazione+deviazione locale) è 18.21 s, corrispondente a 57 mm percorsi a 3.13 mm/s.

La retta LASER, di cui conosciamo l'azimut esatto, a 205.5 parti centesime dell'altezza del foro, cioè 41807 mm, si discosta di 41807 × tan(4'28.8") = 54.5 mm. Restano 2.5 mm di differenza che alla velocità di 3.13 mm/s corrispondono a 0.80 s.

Questi 8 decimi di secondo sono il ritardo che il Sole ha accumulato sulle effemeridi, basate su un ritmo di rotazione terrestre costante, corrisponde al valore cercato di ΔUT1 ed è negativo a causa del ritardo.

Usando ephemvga risultano 0.05 s in più, mentre con le effemeridi della NASA 0.21 s in meno, le differenze dipendono dalle approssimazioni del moto solare adottate nelle varie effemeridi.

Il valore del ritardo di UT1 rispetto ad UTC per il 3 dicembre 2008 pubblicato dall'IERS[20] è di ΔUT1=-0.564 s, compatibile con i risultati trovati in basilica con il metodo qui esposto.

5. *Conclusioni*

È molto importante notare che la precisione raggiungibile nelle osservazioni allo Gnomone Clementino, una volta effettuate precise calibrazioni dello strumento, porti a considerare anche le differenze tra le effemeridi a livello dei centesimi di secondo.

Con il metodo dei transiti paralleli, che poi è una semplice applicazione allo Gnomone Clementino delle tecniche osservative che venivano applicate già all'Osservatorio del Campidoglio proprio per la misura del diametro del Sole[21] e l'uso di video ad alta frequenza temporale, si è giunti a precisioni dell'ordine del decimo di secondo sulla determinazione degli istanti dei transiti meridiani. Questo sorpassa di almeno un ordine di grandezza la stima sull'accuratezza di tali misure allo Gnomone Clementino ferma ancora a 1 secondo.

Le tecniche di ripresa video sviluppate in basilica, sincronizzate con il tempo universale coordinato, sono state esportate in occasione delle missioni osservative delle eclissi totali e anulari di Sole in Spagna, Egitto e Guyana Francese, dedicate alla misura del diametro solare a precisioni del centesimo di secondo d'arco. Queste misure sono poi di supporto alle mis-sioni spaziali solari PICARD[22] e SDO[23] previste per il 2009.

[20] http://www.iers.org/products/13/1388/orig/finals2000A.daily.
[21] Sigismondi (2008b, pp. 183-188).

Lo Gnomone Clementino di Santa Maria degli Angeli è dunque stata una palestra eccezionale per tutti questi studi di astrometria solare di precisione.

Una ricaduta applicativa di questi studi è senz'altro il metodo di orientamento di edifici o strutture ar-cheologiche mediante osservazione dell'istante in cui la luce del Sole è radente ai muri[24]. Se l'orologio è ben sincronizzato, utilizzando le effemeridi calcolate per le coordinate del punto in osservazione, si può ottenere una precisione nell'azimut anche migliore di un minuto d'arco.

*Riferimenti bibliografici*

BEZOARI G., MONTI C., SELVINI A. (2002), *Topografia Generale con Elementi di Geodesia*, UTET, Torino.

BIANCHINI F. (1703), *De Nummo et Gnomone Clementino*, stamperia typis Aloysii, & Francisci de Comitibus, Romae.

BÒNOLI F., PARMEGGIANI G., POPPI F. (2006), *Atti del Convegno Il Sole nella Chiesa: Cassini e le grandi meridiane come strumenti di indagine scientifica*, Giornale di Astronomia 32, Istituti Editoriali e Poligrafici Internazionali, Pisa-Roma.

CATAMO M., LUCARINI C. (2002), *Il Cielo in Basilica, la Meridiana della Basilica di Santa Maria degli Angeli e dei Martiri in Roma*, Ed. A.R.P.A. AGAMI, Roma.

FERRARI DA PASSANO C., MONTI C., MUSSIO L. (1977), *La Meridiana Solare del Duomo di Milano, Verifica e Ripristino nell'anno 1976*, Veneranda Fabbrica del Duomo di Milano, Milano.

HEILBRON J.L. (2005), *Il Sole nelle Chiese. Le grandi Chiese come Osservatori Astronomici*, Editrice Compositori, Bologna.

MONACO G. (1990), *L'Astronomia a Roma, dalle origini al Novecento*, Osservatorio Astronomico di Roma, Roma.

PALTRINIERI G. (2001), *La Meridiana della Basilica di San Petronio in Bologna*, Centro Editoriale Santo Stefano, Bologna.

PAOLI G. (1988), *Ruggero G. Boscovich nella scienza e nella storia del '700*, Accademia Nazionale delle Scienze, Roma.

SIGISMONDI C. (2006), *Pinhole Solar Monitor Tests in the Basilica of Santa Maria degli Angeli in Rome*, in V. Bothmer – A. Hady (eds.), *Solar Activity and its Magnetic Origin*, Proc. IAUC233 Cairo, March 31-April 4 2006, Cambridge University Press, Cambridge, pp. 521-522.

[22] SIGISMONDI – BIANDA – ARNAUD (2008, pp. 189-198).
[23] http://sdo.gsfc.nasa.gov/.
[24] SIGISMONDI (2008c).

Sigismondi C. (2008a), *Stellar and Solar Positions in 1701-3 observed by F. Bianchini at the Clementine Meridian Line*, in R. Ruffini – R.T. Jantzen – H. Kleinert (eds.), *Proc. XI Marcel Grossmann Meeting on General Relativity*, Berlin, July 22-29 2006, World Scientific Publisher, Singapore, pp. 571-573.

Sigismondi C. (2008b), *Measures of Solar Diameter with Eclipses: Data Analysis, Problems and Perspectives*, «American Institute of Physics Conference Proceedings», 1059, pp. 183-188.

Sigismondi C. (2008c), *Effemeridi, Introduzione al Calcolo Astronomico*, Ateneo Pontificio Regina Apostolorum, Roma.

Sigismondi C., Bianda, M., Arnaud J. (2008), *European Projects of Solar Diameter Monitoring*, «American Institute of Physics Conference Proceedings», 1059, pp. 189-198.

US Naval Observatory HM Nautical Almanac Office (2008), *The Astronomical Almanac for the year 2009*, Washington-London.